# Light bullets in optical tandems

Lluis Torner and Yaroslav V. Kartashov

*ICFO-Institut de Ciencies Fotoniques, Mediterranean Technology Park, and Universitat Politecnica de Catalunya, 08860 Castelldefels (Barcelona), Spain*

We address the concept of three-dimensional light bullet formation in structures where nonlinearity and dispersion are contributed by different materials, including metamaterials, which are used at their best to create suitable conditions where bullets can form. The particular geometry considered here consists of alternating rings made of highly dispersive but weakly nonlinear media and strongly nonlinear but weakly dispersive media. We show that light bullets form for a wide range of parameters.

The formation of fully three-dimensional self-trapped wavepackets, or light bullets, is one of the most exciting, yet experimentally unsolved, problems in Optics [1,2]. Light bullets are spatio-temporal solitons that form when a suitable nonlinearity arrests both spatial diffraction and temporal group-velocity dispersion. In principle, bullets may be supported by a variety of nonlinear mechanisms, even though potential solutions tend to be highly prone to dynamical instabilities. Their experimental realization faces two central challenges. Namely, elucidation of a nonlinearity mechanism that supports light bullets as stable entities, and realization of a physical setting where nonlinearity, diffraction and dispersion are all present with suitable strengths without introducing too high propagation losses. A variety of strategies have been found that solve the first challenge. These include quadratic nonlinear media that support solitons for all physical dimensions [3] and where two-dimensional bullet formation was achieved by generating dispersion via achromatic-phase-matching at the expense of one spatial dimension [4,5]; saturable [6] and nonlocal [7,8] media; materials with competing nonlinearities [9,10]; settings where higher-order effects, such as fourth-order dispersion may play a stabilizing role [11]; propagation in optical lattices [12-14]; or filamentation [15,16], just to name a few. However, to date, the second challenge remains unsolved.

A potential approach to overcome this limitation is based on the concept of engineered structures composed of different materials featuring either strong nonlinearity or strong suitable group-velocity dispersion, but not necessarily both together at a given wavelength. Thus, each material is to be used at its best for the purpose at hand. Implementation of such strategy along the longitudinal direction showed that light bullet formation is possible for significantly large tandem domains in the case of quadratic solitons [17]. In this Letter we put forward the concept that stable bullets do form in transverse radially-periodic structures consisting of alternating rings made of a highly dispersive linear material and rings made of a strongly nonlinear material. We find that bullet stability depends crucially on whether the central domain is linear or nonlinear. Here we address materials with cubic saturable nonlinearity, but the concept is expected to hold for different nonlinearities.

We address the propagation of a light beam along the $\xi$ axis of a radial tandem described by the nonlinear Schrödinger equation for dimensionless field amplitude $q(\eta,\zeta,\tau,\xi)$:

$$i\frac{\partial q}{\partial \xi} = -\frac{1}{2}\left(\frac{\partial^2 q}{\partial \eta^2} + \frac{\partial^2 q}{\partial \zeta^2}\right) + \frac{\beta(\eta,\zeta)}{2}\frac{\partial^2 q}{\partial \tau^2} + \sigma(\eta,\zeta)\frac{q|q|^2}{1+S|q|^2}. \qquad (1)$$

Here $\xi = z/L_{\text{difr}}$ is the propagation distance normalized to the diffraction length $L_{\text{difr}} = k_0 r_0^2$; $\eta = x/r_0$ and $\zeta = y/r_0$ are transverse coordinates normalized to the characteristic transverse scale $r_0$; $\tau = t/t_0$ is the normalized (retarded) time; $\beta = L_{\text{difr}}/L_{\text{disp}}$; the dispersion length in each domain is defined as $L_{\text{disp}} = \left|\partial^2 k_0 / \partial \omega^2\right|^{-1} t_0^2$; the quantity $\sigma(\eta,\zeta) = -1$ stands for focusing nonlinearity, while $\sigma(\eta,\zeta) = 0$ stands for a linear medium; $S$ is a saturation parameter. We assume a radially-symmetric structure composed of periodically alternating rings exhibiting anomalous dispersion and weak nonlinearity, where $\beta = -2$ and $\sigma = 0$, and weakly dispersive, but highly nonlinear domains, where $\beta = -0.1$ and $\sigma = -1$. The radial width of the domains is $d$. Importantly, two types of geometries are possible, when the central domain exhibits nonlinearity and when it does not. The refractive index is set to be similar in all domains, thus the structure may be viewed as a *nonlinear lattice*. Note that two-dimensional solitons in geometries where nonlinearity and dispersion are transversally modulated at equal points have been shown to exist [18,19]. The essential ingredient of the approach put forward here, where nonlinearity and dispersion are strong *at different points* of the structure, should be properly appreciated.

Figure 1 illustrates the linear patterns obtained for different radial periods of the tandem in the absence of nonlinearity for Gaussian inputs, i.e., $q|_{\xi=0} = A\exp(-\eta^2 - \zeta^2 - \tau^2)$. For large domain widths diffraction resembles that in uniform media. When $\beta = -2$ in the central domain, dispersion is stronger than diffraction and the intensity isosurfaces have ellipsoidal shapes elongated in time [Fig. 1(a)], while when $\beta = -0.1$ one gets Fig. 1(b). Decreasing the domain width results in distortions of the input since it covers several domains with substantially different dispersions [Fig. 1(c)]. When the domain width is sufficiently small, the beam experiences the average dispersion of the structure. The corresponding intensity isosurfaces approach spheres [Fig. 1(d)]. Thus, addition of nonlinearity may result in the formation of light bullets by compensating diffraction and the effective dispersion.

We search for light bullet solutions in the form $q = w(r,\tau)\exp(ib\xi)$, where $b$ is the propagation constant. Solutions approach those of uniform media with average dispersion and nonlinearity only at the limit $d \to 0$, but in general they rather correspond to the exact nonlinear radial lattice defined by Eq. (1). Here we assume nonlinearity saturation to avoid collapse that occurs in Kerr media. To conduct stability analysis, the perturbed solutions of Eq. (1) can be written as $q = [w + u\exp(ik\phi) + v^*\exp(-ik\phi)]\exp(ib\xi)$, where $u, v$ are small perturbations that can grow with rate $\delta$ upon propagation and $k$ is an azimuthal perturbation index. Typical shapes of light bullets supported by radial tandems are shown in Fig. 2. For suitable parameters, bullets may cover several radial domains and they may feature pronounced shape modulations. The intensity distribution features a ring-like shape in the $(\eta,\zeta)$ plane, an effect that is most pronounced for tandems with a nonlinear central domain. Bullets expand substantially at low and high amplitudes, the latter being a consequence of the nonlinearity saturation. At certain $b$ they acquire minimal width that increases with $S$. The total energy carried by the bullets $U = 2\pi \int_0^\infty r\,dr \int_{-\infty}^\infty w^2 d\tau$ is a non-monotonic function of $b$ [Fig. 3(a)] - it diverges in an upper cutoff $b_{\text{upp}}$ that grows with decreasing $S$ and that only slightly depends on the domain width $d$. When $b$ decreases, the bullet energy reaches a minimal value close to a lower cutoff and then starts increasing as $b \to 0$. Thus, one concludes that in the nonlinear lattice defined by the radial tandems, light bullets always exist above a threshold energy that diminishes with increasing the domain width $d$ and with decreasing the saturation parameter $S$. Figure 3(b) illustrates how the energy carried by the bullets evolves with decreasing the radial domain width $d$. The difference between the corresponding $U(b)$ curves diminishes for any $b$ value (and not only at $b \to 0$ when bullets are always broad and cover many domains) as the domain width $d$ becomes smaller. This is consistent with the expectation that light evolution in the structure with sufficiently small domains mimics the evolution in uniform media with average parameters. However, note that the central result of this Letter is that fully three-dimensional light bullets do exist far from those averaged conditions.

A rigorous linear stability analysis predicts that there exist two types of harmful perturbations that may destabilize light bullets in the radial tandems. The first type of instability corresponds to the azimuthal perturbation index $k=0$ and it is possible only for $b \to 0$ in the narrow region where the dependence $U(b)$ exhibits a negative slope $dU/db \leq 0$. As shown by Fig. 3(c) the corresponding instability domain quickly shrinks and the growth rate (which we found to be purely real) diminishes with decreasing domain width. Azimuthal instabilities associated to $k=1$ may also appear when solutions acquire a ring-like shape in the $(\eta,\zeta)$ plane. We found that such instability is usually absent in tandems with a linear central domain (i.e., in such tandems bullets are stable almost in the entire existence domain) unless $d$ notably exceeds 1. In contrast, in tandems with a nonlinear central domain the azimuthal instability is much stronger, to an extent that may result in the destabilization of light bullets almost in their entire existence domain. Such strong instability may be present even for domain widths $d \sim 0.2$. Example of instabilities are shown in Fig. 3(d), where the curves $\delta(b)$ for $k=1$ are compared for bullets supported by tandems with linear and nonlinear central domains with $d=1.2$. While stabilization in tandems with a nonlinear central domain is possible, the stability region in terms of $b$ is narrow (upper cutoff $b_{\text{upp}} \approx 1.22$) and stabilization occurs at much higher powers than in tandems with a linear central domain. Therefore, instability suppression for light bullets supported by tandems with a nonlinear central core requires substantially smaller domain widths and higher nonlinearity saturation strengths. The predictions by the stability analysis are in agreement with full numerical propagation of perturbed solutions. For example, Fig. 4(a) illustrates how a typical light bullet supported by a tandem featuring a linear central domain cleans up itself and keeps its shape upon propagation in a robust way, while a similar stationary solution supported by a tandem with a nonlinear central domain that eventually decays is shown in Fig. 4(b). Note that in the latter case the input signal goes through significant oscillations and shifts (not directly visible in the isosurface plots).

In summary, stable, fully three-dimensional light bullets can form in radial tandem structures made of alternating linear and nonlinear domains with drastically different dispersion and nonlinear properties. The symmetry and type of nonlinearity addressed here ought to be understood as particular examples, the important result being that light bullets form in structures where materials are used at their best to meet the requirements needed to support light bullets in practice. It is our belief that these findings motivate a program devoted to building suitable metamaterial structures.

# References with titles

1. B. A. Malomed, D. Mihalache, F. Wise, and L. Torner, "Spatiotemporal optical solitons," J. Opt. B **7**, R53 (2005).
2. Y. Silberberg, "Collapse of optical pulses," Opt. Lett. **15**, 1282 (1990).
3. A. A. Kanashov and A. M. Rubenchik, "On diffraction and dispersion effect on three wave interaction," Physica D **4**, 122 (1981).
4. P. Di Trapani, D. Caironi, G. Valiulis, A. Dubietis, R. Danielis, A. Piskarskas, "Observation of temporal solitons in second-harmonic generation with tilted pulses," Phys. Rev. Lett. **81**, 570 (1998).
5. X. Liu, L. J. Qian, and F. W. Wise, "Generation of optical spatiotemporal solitons," Phys. Rev. Lett. **82**, 4631 (1999).
6. D. E. Edmundson and R. H. Enns, "Robust bistable light bullets," Opt. Lett. **17**, 586 (1992).
7. O. Bang, W. Krolikowski, J. Wyller, and J. J. Rasmussen, "Collapse arrest and soliton stabilization in nonlocal nonlinear media," Phys. Rev. E **66**, 046619 (2002).
8. D. Mihalache, D. Mazilu, F. Lederer, B. A. Malomed, Y. V. Kartashov, L.-C. Crasovan, and L. Torner, "Three-dimensional spatiotemporal solitons in nonlocal nonlinear media," Phys. Rev. E **73**, 025601(R) (2006).
9. A. Desyatnikov, A. Maimistov, and B. Malomed, "Three-dimensional spinning solitons in dispersive media with the cubic-quintic nonlinearity," Phys. Rev. E **61**, 3107 (2000).
10. D. Mihalache, D. Mazilu, L.-C. Crasovan, I. Towers, A.V. Buryak, B. A. Malomed, L. Torner, J. P. Torres, and F. Lederer, "Stable spinning optical solitons in three dimensions," Phys. Rev. Lett. **88**, 073902 (2002).
11. G. Fibich and B. Ilan, "Optical light bullets in pure Kerr medium," Opt. Lett. **29**, 887 (2004).
12. B. B. Baizakov, B. A. Malomed, and M. Salerno, "Multidimensional solitons in a low-dimensional periodic potential," Phys. Rev. A **70**, 053613 (2004).

13. D. Mihalache, D. Mazilu, F. Lederer, Y. V. Kartashov, L.-C. Crasovan, and L. Torner, "Stable three-dimensional spatiotemporal solitons in a two-dimensional photonic lattice," Phys. Rev. E **70**, 055603(R) (2004).
14. D. Mihalache, D. Mazilu, F. Lederer, B. A. Malomed, Y.V. Kartashov, L.-C. Crasovan, and L. Torner, "Stable spatiotemporal solitons in Bessel optical lattices," Phys. Rev. Lett. **95**, 023902 (2005).
15. A. Couairon and A. Mysyrowicz, "Femtosecond filamentation in transparent media," Phys. Rep. **441**, 47 (2007).
16. L. Berge and S. Skupin, "Few-cycle light bullets created by femtosecond filaments," Phys. Rev. Lett. **100**, 113902 (2008).
17. L. Torner, S. Carrasco, J. P. Torres, L.-C. Crasovan, and D. Mihalache, "Tandem light bullets," Opt. Commun. **199**, 277 (2001).
18. H. Sakaguchi and B. A. Malomed, "Two-dimensional solitons in the Gross-Pitaevskii equation with spatially modulated nonlinearity," Phys. Rev. E **73**, 026601 (2006).
19. H. Sakaguchi and B. A. Malomed, "Channel-guided light bullets," Phys. Rev. A **75**, 063825 (2007).

# References without titles

1. B. A. Malomed, D. Mihalache, F. Wise, and L. Torner, J. Opt. B **7**, R53 (2005).
2. Y. Silberberg, Opt. Lett. **15**, 1282 (1990).
3. A. A. Kanashov and A. M. Rubenchik, Physica D **4**, 122 (1981).
4. P. Di Trapani, D. Caironi, G. Valiulis, A. Dubietis, R. Danielis, A. Piskarskas, Phys. Rev. Lett. **81**, 570 (1998).
5. X. Liu, L. J. Qian, and F. W. Wise, Phys. Rev. Lett. **82**, 4631 (1999).
6. D. E. Edmundson and R. H. Enns, Opt. Lett. **17**, 586 (1992).
7. O. Bang, W. Krolikowski, J. Wyller, and J. J. Rasmussen, Phys. Rev. E **66**, 046619 (2002).
8. D. Mihalache, D. Mazilu, F. Lederer, B. A. Malomed, Y. V. Kartashov, L.-C. Crasovan, and L. Torner, Phys. Rev. E **73**, 025601(R) (2006).
9. A. Desyatnikov, A. Maimistov, and B. Malomed, Phys. Rev. E **61**, 3107 (2000).
10. D. Mihalache, D. Mazilu, L.-C. Crasovan, I. Towers, A.V. Buryak, B. A. Malomed, L. Torner, J. P. Torres, and F. Lederer, Phys. Rev. Lett. **88**, 073902 (2002).
11. G. Fibich and B. Ilan, Opt. Lett. **29**, 887 (2004).
12. B. B. Baizakov, B. A. Malomed, and M. Salerno, Phys. Rev. A **70**, 053613 (2004).
13. D. Mihalache, D. Mazilu, F. Lederer, Y. V. Kartashov, L.-C. Crasovan, and L. Torner, Phys. Rev. E **70**, 055603(R) (2004).
14. D. Mihalache, D. Mazilu, F. Lederer, B. A. Malomed, Y.V. Kartashov, L.-C. Crasovan, and L. Torner, Phys. Rev. Lett. **95**, 023902 (2005).
15. A. Couairon and A. Mysyrowicz, Phys. Rep. **441**, 47 (2007).
16. L. Berge and S. Skupin, Phys. Rev. Lett. **100**, 113902 (2008).
17. L. Torner, S. Carrasco, J. P. Torres, L.-C. Crasovan, and D. Mihalache, Opt. Commun. **199**, 277 (2001).
18. H. Sakaguchi and B. A. Malomed, Phys. Rev. E **73**, 026601 (2006).
19. H. Sakaguchi and B. A. Malomed, Phys. Rev. A **75**, 063825 (2007).

# Figure captions

Figure 1. Isosurface plots showing output patterns after propagation at $\xi = 2$ of input Gaussian signals in linear tandem structures with (a), (b) $d = 20$, (c) $d = 0.8$, and (d) $d = 0.4$. In (a) the dispersion in the central domain is set to $\beta = -2$. In (b), (c) and (d), $\beta = -0.1$.

Figure 2. Soliton profiles in tandem structures with $d = 0.4$ and $S = 0.5$. Panels (a), (c) correspond to $b = 0.3$, while panels (b),(d) correspond to $b = 0.85$. In (a), (b) central domain is linear, while in (c),(d) central domain is nonlinear.

Figure 3. Energy versus propagation constant for (a) different $S$ values at $d = 0.4$ and (b) different $d$ values at $S = 0.5$. Circles in (a) correspond to solitons shown in Figs. 2(a) and 2(b). In (b) domain width takes values $d = 0.8$, $0.6$, $0.4$, $0.2$, and $0.1$ from lower to upper curve. (c) Perturbation growth rate versus $b$ at $k = 0$, $S = 0.5$. In panels (a)-(c) central domain is linear. (d) Growth rate versus $b$ at $k = 1$, $d = 1.2$, $S = 0.5$ in the structure with linear (1) and nonlinear (2) central domains.

Figure 4. (a) Isosurface plots at $\xi = 0$ (left) and $\xi = 128$ (right) showing stable propagation of light bullet in tandem structure with linear central domain with $\beta = -2$, and (b) isosurface plots at $\xi = 0$ (left) and $\xi = 72$ (right) showing unstable propagation of light bullet in tandem with nonlinear central domain with $\beta = -0.1$. In all cases $b = 0.3$, $d = 0.4$, $S = 0.5$.

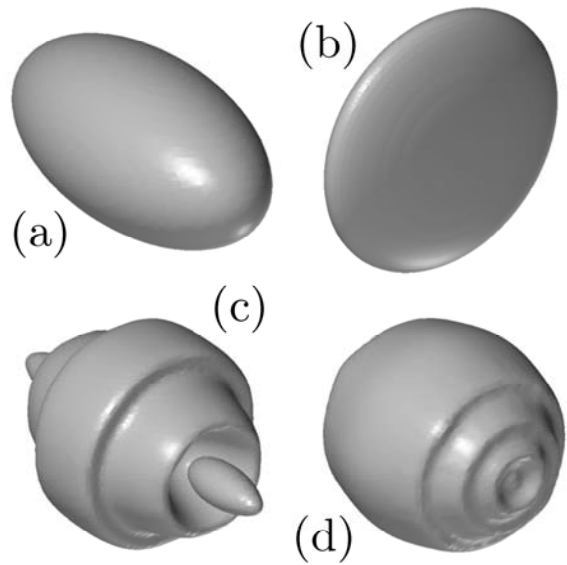

Figure 1. Isosurface plots showing output patterns after propagation at $\xi = 2$ of input Gaussian signals in linear tandem structures with (a), (b) $d = 20$, (c) $d = 0.8$, and (d) $d = 0.4$. In (a) the dispersion in the central domain is set to $\beta = -2$. In (b), (c) and (d), $\beta = -0.1$.

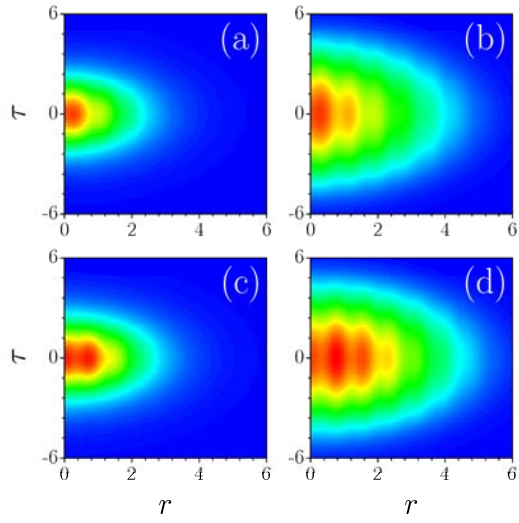

Figure 2.  Soliton profiles in tandem structures with $d = 0.4$ and $S = 0.5$. Panels (a), (c) correspond to $b = 0.3$, while panels (b),(d) correspond to $b = 0.85$. In (a), (b) central domain is linear, while in (c),(d) central domain is nonlinear.

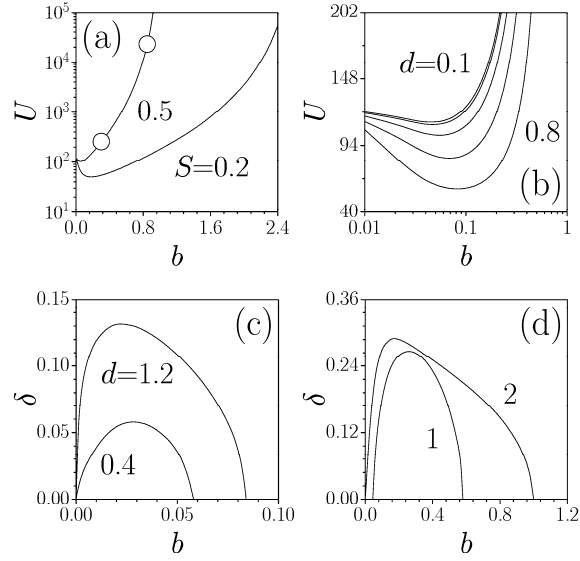

Figure 3. Energy versus propagation constant for (a) different $S$ values at $d = 0.4$ and (b) different $d$ values at $S = 0.5$. Circles in (a) correspond to solitons shown in Figs. 2(a) and 2(b). In (b) domain width takes values $d = 0.8$, $0.6$, $0.4$, $0.2$, and $0.1$ from lower to upper curve. (c) Perturbation growth rate versus $b$ at $k = 0$, $S = 0.5$. In panels (a)-(c) central domain is linear. (d) Growth rate versus $b$ at $k = 1$, $d = 1.2$, $S = 0.5$ in the structure with linear (1) and nonlinear (2) central domains.

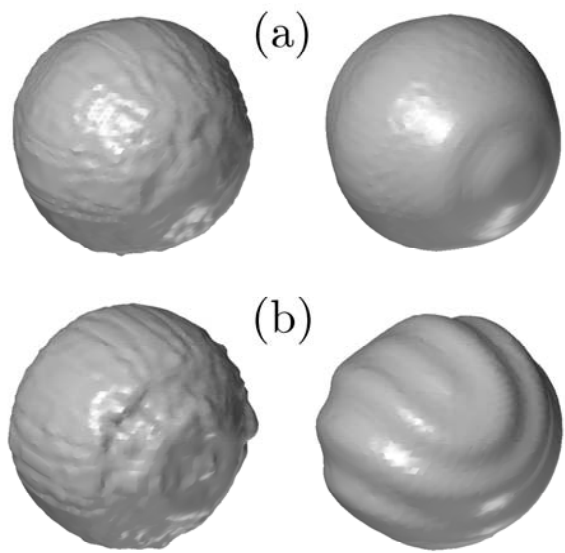

Figure 4. (a) Isosurface plots at $\xi = 0$ (left) and $\xi = 128$ (right) showing stable propagation of light bullet in tandem structure with linear central domain with $\beta = -2$, and (b) isosurface plots at $\xi = 0$ (left) and $\xi = 72$ (right) showing unstable propagation of light bullet in tandem with nonlinear central domain with $\beta = -0.1$. In all cases $b = 0.3$, $d = 0.4$, $S = 0.5$.